# Detecting single viruses and nanoparticles using whispering gallery microlasers

Lina He, Sahin Kaya Ozdemir, Jiangang Zhu, Woosung Kim, Lan Yang[*]
Department of Electrical and Systems Engineering, Washington University in St. Louis, MO 63130, USA
* yang@ese.wustl.edu

**Detection and characterization of individual nano-scale particles, virions, and pathogens are of paramount importance to human health, homeland security, diagnostic and environmental monitoring[1]. There is a strong demand for high-resolution, portable, and cost-effective systems to make label-free detection and measurement of individual nanoparticles, molecules, and viruses[2-6]. Here, we report an easily accessible, real-time and label-free detection method with single nanoparticle resolution that surpasses detection limit of existing micro- and nano-photonic devices. This is achieved by using an ultra-narrow linewidth whispering gallery microlaser, whose lasing line undergoes frequency splitting upon the binding of individual nano-objects. We demonstrate detection of polystyrene and gold nanoparticles as small as 15 nm and 10 nm in radius, respectively, and Influenza A virions by monitoring changes in self-heterodyning beat note of the split lasing modes. Experiments are performed in both air and aqueous environment. The built-in self-heterodyne interferometric method achieved in a microlaser provides a self-reference scheme with extraordinary sensitivity[7,8], and paves the way for detection and spectroscopy of nano-scale objects using micro- and nano-lasers.**

Most of the biological agents and synthetic particles of interest have low polarizability due to their small size and low refractive index contrast with the surrounding medium, leading to weak light-particle interactions which make their label-free optical detection at single particle resolution difficult. Micro- and nano-photonic resonant devices have emerged as highly sensitive platforms for detection of individual virions and nanoparticles due to the significantly enhanced light-matter interactions originating from the high ratio of their quality-factor ($Q$) to mode volume ($V$)[7,9-15]. For example, detection of single Influenza virions and polystyrene nanospheres as small as 30 nm in radius have been demonstrated in whispering gallery mode (WGM) microspheres using reactive shift[12] and in microtoroids using mode splitting technique[7], respectively. In both techniques, wavelength of a tunable laser is scanned to obtain the transmission spectrum of a resonant mode from which specific information, either the resonance shift and/or the mode splitting, is extracted to detect and measure nanoparticles. The ultimate detection limit strongly relies on $Q/V$ which not only determines the light-matter interaction strength but also sets the smallest resolvable changes in the WGM spectrum[16]. Higher $Q$, limited by material absorption, implies narrower resonance linewidth and better resolution.

Here we report the first microlaser-based detection scheme with single particle resolution surpassing the detection capabilities of existing micro- and nano-photonic devices. The ultimate detection limit is set by the laser linewidth, which can be as narrow as a few Hertz for a WGM microlaser, and is certainly much narrower than the resonance linewidth of any passive resonators[17]. Thus, a WGM microlaser sensor, regardless of whether it utilizes reactive shift[12,18] or mode splitting [7], has the potential to detect smaller objects beyond the reach of passive resonator sensors.

A WGM microlaser supports two frequency-degenerate but counter-propagating lasing modes, which are confined in the resonator with evanescent tails probing the surrounding medium many times. A particle that enters the evanescent field of the microlaser couples the degenerate lasing modes to each other via intracavity Rayleigh back-scattering, and thus leads to laser frequency splitting. Spectral distance between the split lasing modes is determined by particle polarizability and particle position in the mode volume[19]. The polarizability of a spherical particle of radius $R$ is given by $\alpha = 4\pi R^3 (n_p^2 - n_s^2) / (n_p^2 + 2n_s^2)$, where $n_p$ and $n_s$ are the refractive indices of the particle and the surrounding medium, respectively. Thus, any change in $\alpha$ is translated into a change in frequency splitting. Similarly, any particle binding event inducing excess polarizability is observed as a change in the splitting[7,20]. A simple and cost-effective method of measuring the frequency splitting is to mix the split laser modes in a photodetector, and thus the output is a heterodyne beat note with beat frequency corresponding to the frequency splitting[21]. Therefore, nanoparticle adsorption events can be revealed in real time by monitoring the beat note.

In our experiments performed in air, the WGM microlasers have toroidal-shaped structure fabricated from Erbium (Er)-doped silica[22,23]. The concentration of $Er^{3+}$ ions in silica is $5\times10^{18}$ ions/cm$^3$, which assures continuous-wave laser operation. The resonators are 20-40 μm in diameter, and have $Q \sim 6\times10^6$. Experimental setup is shown in Fig. 1a. The fiber taper coupled microlaser is continuously pumped by a light source at around 1.46 μm within the $Er^{3+}$ absorption band. Laser emission in 1.55 μm band is monitored by a photodiode connected to an oscilloscope. Figure 1b depicts the working principle of nanoparticle detection. Single particle binding events are translated into discrete frequency changes of the heterodyne beat signal (Supplementary Fig. S10). Thus, particle detection and counting are achieved by monitoring the beat frequency.

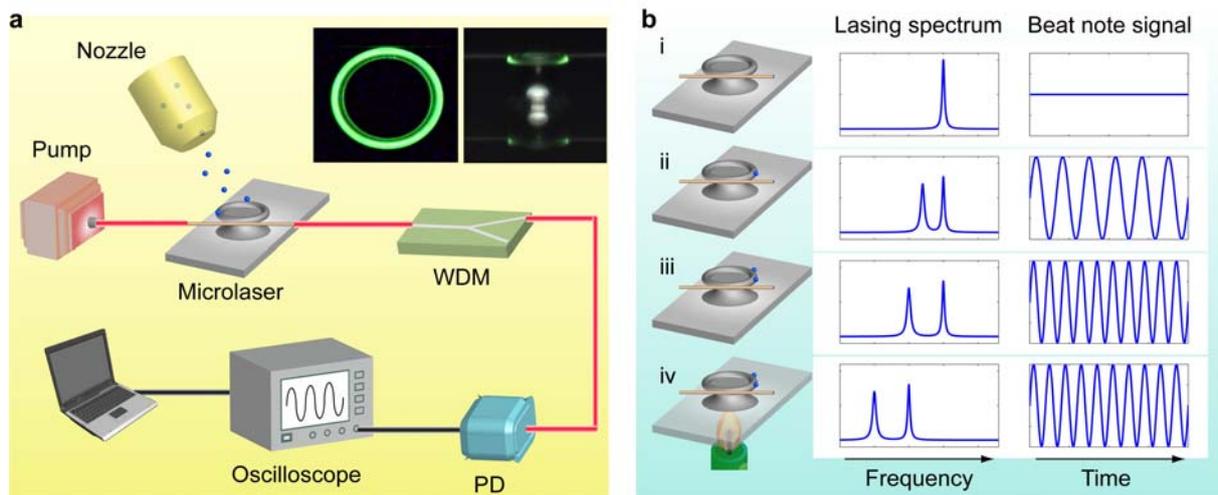

**Figure 1. Heterodyne detection of single nano-objects using frequency splitting in a microlaser. a**, Schematic illustration of the experimental setup. A nozzle continuously delivers viral or synthetic nanoparticles onto a toroidal-shaped microlaser[7], which translates changes in polarizability into changes in frequency splitting. The pump light and the split lasing modes are separated using a wavelength division multiplexer (WDM). Split lasing modes are mixed in a photodetector (PD) leading to a heterodyne beat note signal. Inset: Experimentally observed green ring due to up-conversion of $Er^{3+}$ ions traces the WGM along the periphery of the microresonator (top and side views). **b**, Before the arrival of nanoparticles, there is a single laser mode with constant laser intensity (i). With the arrival of the first nanoparticle, lasing mode splits into two leading to a beat note whose frequency corresponds to the frequency splitting (ii). Subsequent particle binding event changes the beat frequency (iii). Since the split lasing modes reside in the same microlaser, environmental noise such as temperature fluctuations (illustrated by a heat source placed under the chip) affects both modes in the same way leading to a self-referencing scheme (iv). Thus, although each split mode undergoes spectral shift, the splitting between them does not change, making this detection scheme resistant to environmental noises[8].

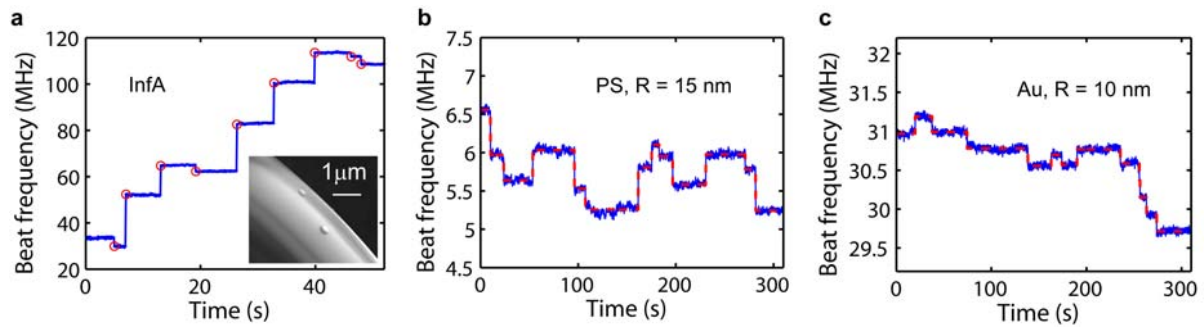

**Figure 2. Discrete changes in beat frequency of split lasing modes in response to the arrival of nano-objects onto the microlaser continuously. a**, InfA virions. Inset in (**a**) is the scanning electron microscopy image of two InfA virions on a microtoroidal laser. **b**, PS nanoparticles with radius $R$ = 15 nm. **c**, Au nanoparticles with radius $R$ = 10 nm. Each discrete jump corresponds to the adsorption of a single nano-object. Red circles in (**a**) denote individual virus binding events. Red lines in (**b**) and (**c**) are drawn at mean of the measured beat frequencies for each binding event.

We challenged the single particle detection capability of the microlaser using polystyrene (PS) nanoparticles of mean radius $R$ = 15 nm, gold (Au) nanoparticles of $R$ = 10 nm (CV<8%, CV: coefficient of variation, defined as the ratio of standard deviation to mean), and Influenza A (InfA) virions. Figure 2 shows the evolution of beat frequency as particles enter the mode volume of a microlaser one by one. Each discrete upward or downward jump in the beat frequency indicates a single particle adsorption event. The directions and heights of the discrete jumps are related to the polarizability of each arriving particle and to the location of each particle with respect to the previously adsorbed particles (Supplementary Fig. S1). Due to the trapping force from the resonator field, desorption of nanoparticles from the resonator happens rarely. Detection limits for single PS and Au nanoparticles shown in Figs. 2b and 2c using the proposed scheme surpass the reported detection limits for passive microresonators[7,12,13].

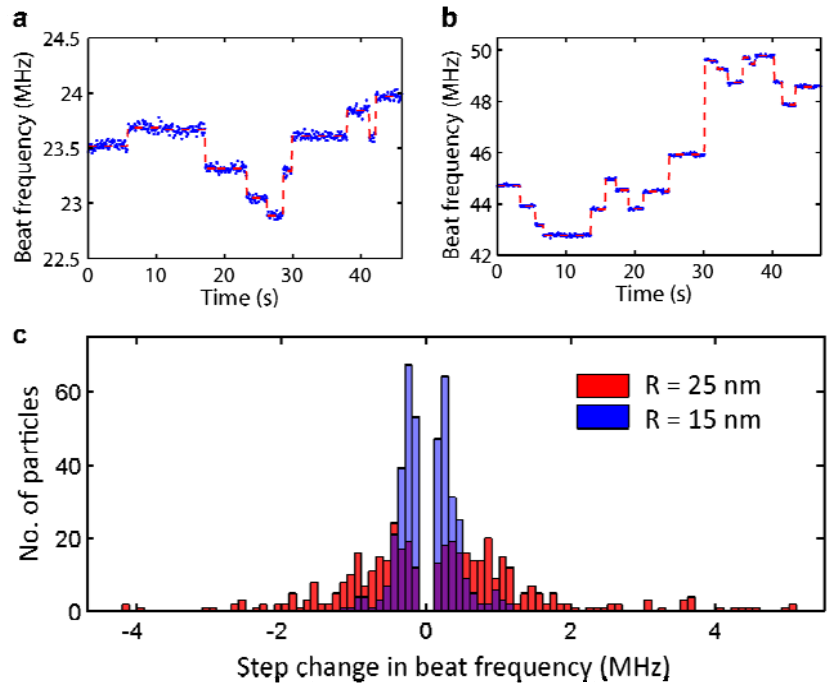

**Figure 3. Estimating particle size with ensemble measurement.** Real time records of beat frequency when Au nanoparticles of radius $R$ = 15 nm (**a**) and $R$ = 25 nm (**b**) are deposited one by one onto the microlaser randomly. Red lines represent mean of the measured beat frequencies for each binding event. **c,** Histograms of the heights of discrete jumps in beat frequency for two different Au particle sizes. Binding events with step heights in the range ±100 kHz are rejected as they are biased with the beat frequency fluctuation noise. The total numbers of detected binding events are 397 for $R$ = 15 nm and 419 for $R$ = 25 nm. The measured standard deviations of the histograms are 0.383 MHz for $R$ = 15 nm and 1.344 MHz for $R$ = 25 nm. All the data, for a total number of 816 Au nanoparticles, is measured using the same microlaser and the same lasing mode, implying that small nanoparticles do not cause significant change in the linewidth of the lasing modes.

Figures 3a and 3b show the heterodyne beat frequencies obtained from ensemble measurement of Au particles of radius $R$ = 15 nm (CV<8%) and $R$ = 25 nm (CV<8%), respectively. Particles are deposited one by one at random locations on the microlaser surface. All measurements are performed using the same microlaser and the same lasing mode to minimize the device and mode related effects. The constructed histograms for the heights of jumps in beat frequency clearly show that the standard deviation for $R = 25$ nm particles is larger than that for $R = 15$ nm particles (Fig. 3c). This information can be used to extract the polarizability and thus the size of unknown particles by using some reference particles (Supplementary Figs. S11 and S12).

Linewidth broadening of the lasing modes due to losses induced by nanoparticles of $R < 250$ nm is much smaller than the induced frequency splitting (Supplementary Fig. S2). Therefore, a large number of particles can be detected using the same lasing mode in a single microlaser without significantly degrading the lasing linewidth. In the experiments, we detected and counted up to 816 continuously deposited Au nanoparticles using the same mode in a microlaser (Fig. 3c).

To demonstrate simultaneous multi-wavelength detection capability of the proposed scheme, we used a microlaser supporting two lasing modes (Fig. 4a). The same pump source was utilized to excite the two lasing modes. In this case, the particle-induced frequency splitting in each mode is determined by its spatial overlap with the nanoparticle, and thus each lasing mode produces its own heterodyne beat note. The detected signal becomes a mixture of these beat notes which are revealed by two frequency peaks in the fast Fourier transform (FFT) spectrum (Fig. 4b). Figure 4c depicts the evolution of the two beat frequencies as particles are continuously deposited onto the microlaser. From the intensity graph in Fig. 4c, the magnitude of beat frequency is different for the two modes enabling to discriminate and track the two beat notes. At each binding event, the two lasing modes experience different changes. This can be exploited to decrease the false-negative results (e.g., a particle is adsorbed onto the surface but is not detected) as shown in Fig. 4d. Thus, using multiple-wavelength microlasers helps reduce the detection errors induced by the position-dependent change in frequency splitting. These results show that the microlaser provides a compact platform to realize simultaneous multi-wavelength detection of nano-objects.

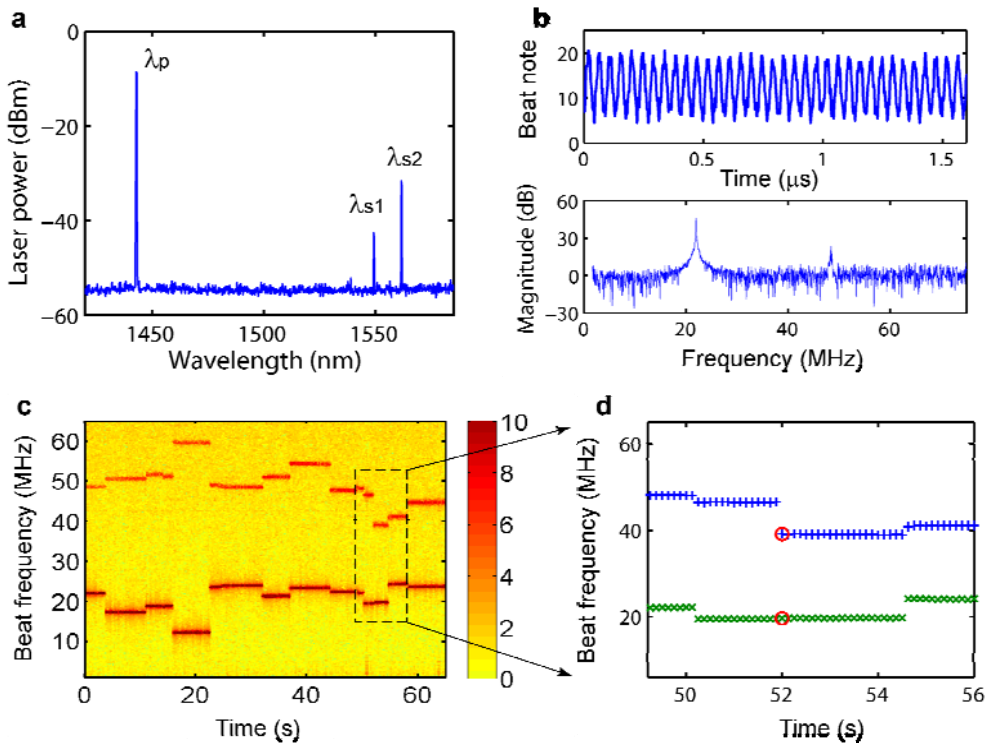

**Figure 4. Simultaneous multi-wavelength detection of nanoparticles using a single microlaser. a**, Typical spectrum of a two-mode microlaser. The pump is located at $\lambda_p$ = 1443 nm, and the two lasing lines are at $\lambda_{s1}$ = 1549 nm and $\lambda_{s2}$ = 1562 nm. **b**, A typical beat note signal and its FFT spectrum for a two-mode microlaser. The two peaks in the FFT spectrum correspond to frequency splitting of the lasing modes at $\lambda_{s1}$ and $\lambda_{s2}$, separately. **c**, Intensity graph of the FFT spectrum when Au particles of $R$ = 50 nm (CV<8%) are continuously deposited onto the microlaser. The side bar denotes magnitude of the FFT spectrum in dB. For the same particle binding events, the heights of discrete jumps in the two beat frequencies are different. **d**, Close-up of the black

rectangle in (**c**). Red circles mark the particle binding event that is clearly detected by one laser mode, but undetected by the other mode.

The unprecedented lower detection limit provided by the microlaser stems from its narrow linewidth and self-referencing nature. Hertz level linewidths have been reported for lasing in Er-doped WGM microcavities[17]. This suggests that detection of frequency splitting as small as a few tens of Hertz, which translates into a lower detection limit less than 1 nm, is within reach of WGM microlasers (Supplementary Figs. S8 and S9). In addition to its superior lower detection limit, a microlaser based detection scheme also eliminates the need for a narrow-linewidth tunable laser to detect the induced spectral shift or mode splitting. Thus, tunable laser related issues such as slow response time, thermal effects, and tuning noises (e.g., piezo-motion) are avoided. In principle, the ultimate response time for passive resonator-based sensors is limited by the cavity lifetime. However, in practice it is limited by the wavelength scanning speed, which for piezo-tuned systems is on the order of milliseconds [7]. For our scheme, the response time is mainly limited by the data acquisition system and can be on the order of microseconds. Thus, high detection sensitivity and real-time measurements are ensured without the need for active stabilization of pump light or temperature.

Although the theoretical detection limit is set by the linewidths of the split lasing modes, the device geometry and the noise level are the main factors in practice. The former is related with the microlaser mode distribution, which determines the mode volume and the evanescent field probing the surrounding medium (Supplementary Figs. S3-S6), and the latter is determined by fluctuations in the beat frequency (Supplementary Fig. S7). We observed in experiments and confirmed with numerical analysis that a smaller microcavity provides a smaller mode volume together with a larger evanescent field in the surrounding, and hence leads to a higher sensitivity. We measured the noise level of beat frequency to be ± 100 kHz which did not change much with the incoming particles. The detection limit in this study was achieved in an operating environment with no stabilization or external reference, and could be further improved by better control and characterization of the spectral properties of the pump laser, by operating the microlaser in a more stable environment, and by employing a better data acquisition system.

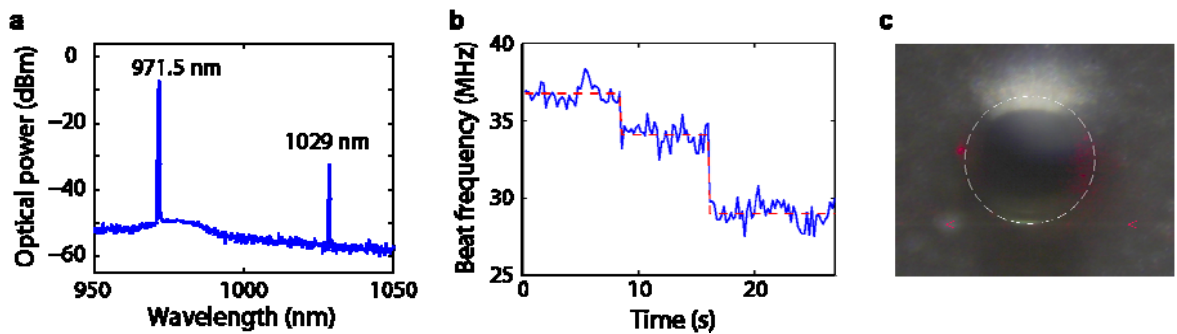

**Figure 5. Detection of nanoparticles in water using frequency splitting in a microlaser. a**, Lasing spectrum of Yb-doped microtoroid laser in water. Laser emission occurs at 1029 nm when pumped at 971.5 nm. **b**, Changes in beat frequency with time after the PS nanoparticle suspension (R = 30 nm and CV = 16.8%) is injected into the chamber. **c**, Top view of the microlaser in water with a particle bound on its ring. The particle becomes visible due to scattering of red light coupled into the microlaser. The white light at the top is due to the illuminating light from the microscope. The white dotted lines are drawn to denote the boundary of the microlaser and the fiber taper. The red arrows show the direction of light propagation in the fiber taper.

Finally, we tested the proposed detection scheme in aqueous environment. The fiber taper coupled microlaser was immersed in a chamber filled with 2 mL of water. A syringe pump connected to the chamber was used to inject PS particles into the chamber[24]. For water experiments, we used Ytterbium (Yb) –doped microlasers due to the low absorption of water at $Yb^{3+}$ emission band of 1040 nm[25]. Figure 5 shows the lasing spectrum of a Yb-microlaser (diameter ~ 100 μm) in water together with the observed discrete jumps in the beat frequency as PS particles bind to the microlaser. This first demonstration of particle-induced frequency splitting in a microlaser placed in aquatic environment completes a crucial step for extending the highly sensitive technique to sensing in aqueous and biological fluids.

In this study, we demonstrated a detection scheme using an ultra-narrow linewidth WGM microlaser which allows detection of nano-scale objects beyond the reach of schemes using passive microcavities. Detection and counting of individual nanoparticles and virions are achieved by monitoring discrete changes in the heterodyne beat frequency produced by the split lasing modes in microlasers. A significant improvement in the limit of label-free detection is demonstrated without increasing the system complexity. Histograms of step changes in frequency splitting can be used to estimate the particle size.

The proposed scheme could be further improved to allow single-shot size measurement by monitoring changes in both frequencies and linewidths of the split lasing modes. This may be done by employing linewidth measurement techniques[26]. The ultra-low threshold and narrow-linewidth microlasers can be integrated with on-chip laser diodes as pump sources enabling a compact measurement platform. Moreover, microlasers with multiple-wavelength emission can be configured to classify nanoparticles[27]. Although we report experiments using microtoroidal cavity lasers, the techniques developed here can be used in many other nano- and micro-lasers, such as photonic crystal cavities and WGM microlasers using microspheres, microrings, etc.

**Acknowledgements**

We gratefully acknowledge the support from NSF under Grant No. 0907467. This work was performed in part at the NRF/NNIN (NSF award No. ECS-0335765) of Washington University in St. Louis. We thank Prof. Chen of EECE at Washington University in St. Louis for providing instrumentation for nanoparticle deposition; and F. Monifi and B. Peng for stimulating discussions.

**Author Contributions**

L.H., S.K.O., J.Z., and L.Y. designed the experimental concept. L.H. and J.Z. performed the experiments in air. L.H. and W.K. performed the experiments in water. L.H. and S.K.O. contributed to the theoretical work. L.Y. supervised the project. All authors contributed to the discussion of the results and the preparation of the manuscript.

**Additional information**

The authors declare no competing financial interests.

Supplementary Information for

Detecting single viruses and nanoparticles using whispering gallery microlasers

Lina He, Sahin Kaya Ozdemir, Jiangang Zhu, Woosung Kim, and Lan Yang[*]
Department of Electrical and Systems Engineering, Washington University in St. Louis, MO 63130, USA
* yang@ese.wustl.edu

**Microlasers for sensing: breaking the detection limit of passive resonators**

In the past few years, there have been tremendous research efforts using whispering gallery mode (WGM)[S1-3] and photonic crystal (PC) resonators[S4-6] for detection and characterization of nanoscale objects. The underlying principle in these resonator-based detection schemes is the change in effective refractive index or effective polarizability of the resonator-surrounding system upon the arrival of a nanoscale object in the resonator mode volume. The change leads to a shift in the resonance frequency and/or the splitting of a single resonant mode into two. Detection method based on tracking the resonant frequency shifts is referred to as *spectral shift method (SSM),* whereas that relies on tracking the changes in the mode splitting is referred to as *mode splitting method (MSM)*.

Despite the great progress shown in the past few years in terms of pushing the detection limit for single particles, there are some fundamental limitations for the SSM and MSM techniques employed in the current state-of-the-art resonators. The fundamental limitations are linked to the well-known resonator parameter $Q/V$, the ratio of the quality factor $Q$ and the mode volume $V$, which is pivotal in determining the interaction strength between the cavity field and the nano-object[S7]. It is crucial that the highest $Q$ and the smallest $V$ are achieved simultaneously to reach the ultimate detection capabilities of the resonator-based schemes. Higher $Q$ implies lower loss, narrower resonance linewidth, and longer photon lifetime in the cavity. Longer photon time allows longer interaction time between the cavity photons and the nano-object. Narrower resonance linewidth indicates smaller detectable changes in the resonance spectrum. Smaller $V$, on the other hand, implies a more confined field, that is a higher power per unit area. Consequently, the nano-object in the mode volume experiences significantly enhanced field intensity, and even an object with small size in nanoscale or small refractive index contrast with the surrounding medium may scatter significant amount of light which will enable its detection. It is clear that the field intensity at the location of the nano-object is important. Even within a very small mode volume $V$, the field is not uniformly distributed, e.g., the field exponentially decays from its maximum value at the resonator boundary in the radial direction of a WGM resonator. The dependence of light-matter interaction on the field distribution could be characterized by the normalized field distribution denoted as $f$. The higher the $f$, the higher the field will be screened by the nano-object. In short, one should minimize $V$ and maximize $f$ and $Q$ to increase the sensitivity and resolvability and thus to reduce the detection limit of the resonator scheme. The effect of $f$, $V$, and $Q$ on the amount of perturbations induced by particles on the resonator system for the case of MSM will be clarified through numerical simulations in the following sections.

Photonic crystal resonators possess extremely small mode volume (on the order of $\lambda^3$, where $\lambda$ is the operation wavelength), and therefore provide a high value of $f^2/V$ thus is very sensitive to perturbations[S6]. However, in such resonators with nanoscale structures, in most cases it is difficult to obtain very high $Q$ factors due to fabrication imperfection and bending and radiation losses. For WGM resonator sensors, on the other hand, $V$ is higher than that of PC resonators. Therefore, WGM sensors provide smaller $f^2/V$ and smaller susceptibility to perturbations than PC resonators. However, $Q$ of WGM resonators is higher than that of PC resonators. For example, $Q$ factor exceeding $10^8$ can be easily obtained in silica microsphere and microtoroid resonators[S8-9]. Due to the ultra-high $Q$ factor, resonance shift of the order of femtometer can be resolved[S2]. To date, the reported smallest particle that was detected and sized using a WGM resonator is 30 nm in radius[S3]. To obtain a better limit with WGM resonators, we need to reduce the mode volume $V$ and/or enhance the $Q$ factor. The mode volume can be reduced by decreasing the cavity size. However, cavity size cannot be made arbitrary small without sacrificing the $Q$ factor. If the size of the resonator becomes smaller than some threshold value, its $Q$ will degrade significantly, because, for smaller resonators, the radiation loss increases and eventually becomes the dominant loss mechanism reducing the $Q$ factor. By calculating the field distribution in microtoroids using finite element method, the major diameter of a silica microtoroid needs to be greater than 22 μm to keep $Q$ factor higher than $10^8$ in 1550 nm band. The ultimate $Q$ factor of a WGM resonator is limited by the cavity material loss, e.g., $Q$ factor of a fused-silica microsphere cannot exceed $10^{10}$ due to material loss[S8,10].

It has been previously discussed and shown that passive resonators can be doped with active material (i.e., gain medium) so that a part of the cavity losses will be compensated by the gain provided by the active material[S11-13]. Although introducing such dopants into material-matrix of a resonator at first reduces the $Q$ factor below that of the un-doped resonator (i.e., termed as the "*passive resonator*"), one can increase the $Q$ of the doped-resonator (i.e., termed as the "*active resonator*") beyond that of the passive resonator by optical or electrical pumping. Emission from the excited gain medium into the resonant modes of interest partially compensates for the resonator loss leading to an overall increase in $Q$ factor. If the resonator is pumped above its lasing threshold, stimulated emission exceeds the spontaneous emission leading to laser oscillation in the resonator, and hence forming a laser. The linewidth of the lasing mode is limited by the spontaneous emission noise and can be as small as a few Hz[S14], which is much narrower than that of a passive resonator. The fundamental linewidth of a laser $\Delta\nu_{laser}$, if technical noise is ignored, is given by the Schawlow-Townes formula[S15] as

$$\Delta\nu_{laser} = \frac{\pi h \nu (\Delta\nu)^2}{P_{out}} \qquad (S1)$$

where $h\nu$ is the photon energy with $h$ representing the Planck's constant and $\nu$ denoting the cavity resonant frequency, $\Delta\nu$ is the cold cavity resonance linewidth (i.e., when gain medium is not excited) that is related to the cold cavity $Q$ through $\Delta\nu = \nu/Q$, and $P_{out}$ is the laser output power. For cold cavity $Q$ in the range $10^6 \sim 10^7$, lasing wavelength $\lambda = 1550$ nm, and output laser power $P_{out} = 10$ μW, the laser linewidth from the cavity is estimated from Eq. (S1) to be in the range of 15 Hz ~ 1.5 kHz. Therefore, a microlaser-based sensing scheme has the potential to provide ultra-low detection limit, beyond the reach of any passive resonator-based sensor.

It should be noted here that microlaser-based approach will significantly contribute to the detection capabilities of both methods, that is, both SSM and MSM will benefit from the ultra-narrow linewidth of microlasers. In SSM, frequency shift of the narrow laser line will be detected in electrical domain by heterodyning with an external local oscillator field with a smaller or comparable linewidth; in MSM, on the other hand, one of the split lasing modes acts as a local oscillator for the other, forming a self-heterodyning detection scheme. A distinct advantage of MSM over SSM is its self-referencing property which originates from the fact that the split lasing modes reside in the same resonator and are affected in the same way by the interfering

perturbations[S16]. Thus the effect of such perturbations can be minimized as one of the split modes acts as a reference to the other. In SSM, detection and measurement are done using the spectral properties of a single lasing mode which makes it difficult to discriminate interfering perturbations from those of interest. As demonstrated with passive resonators, MSM allows single-shot size measurement of nano-objects as differences in the spectral properties (resonance frequency and linewidth) of the split modes carry the polarizability information of the object. One can do the same thing with MSM in a microlaser provided that changes in the linewidth and frequency of the split lasing modes are accurately monitored. In SSM, on the other hand, one should perform statistical techniques on the data obtained from ensemble measurements, i.e., single-shot size measurement cannot be performed using SSM.

**Theory of frequency splitting induced by multiple-particle binding on a resonator**

In optical microcavities, each traveling WGM possesses a twofold degeneracy due to two propagation directions: clockwise (CW) and counterclockwise (CCW). These two degenerate modes have the same resonance frequency and linewidth. When sub-wavelength scatterers in the optical path couple a portion of the energy of one of the modes into the other mode propagating in the opposite direction, this degeneracy could be lifted leading to mode splitting. Superposition of these two traveling-wave modes forms two orthogonal standing-wave modes (SWMs) within the resonator[S17-18]. Field distributions of these two SWMs have a spatial phase difference of $\pi/2$, i.e., the field node of one of the SWMs corresponds to the anti-node of the other. The two SWM fields have different overlaps with the scatterers, and thus their spectral properties differ, which are reflected as differences in the resonance frequency shift and resonance linewidth broadening they experience with respect to the original degenerate mode. This effect is responsible for the observed frequency splitting.

Frequency difference of the two SWMs is determined by the coupling strength between CW and CCW modes, whereas the linewidth difference depends on the coupling of CW and CCW modes to the environment and to other modes which open additional decay channels[S19-20].

In the case of one single scatterer on the resonator, one SWM places its node at the location of the scatterer while the other has its anti-node at the same location. The frequency splitting and linewidth difference of the two SWMs are characterized by $2g = -\alpha f^2(r) \omega_c / V$ and $2\Gamma = \alpha^2 f^2(r) \omega_c^4 / (3\pi v^3 V)$, respectively, where $\alpha$ is the polarizability of the scatterer, $f(r)$ represents the normalized (i.e., normalized to the maximum value) distribution of the WGM field magnitude at the location $r$ of the scatterer, $\omega_c$ is the resonance angular frequency, and $V$ is the WGM mode volume. It is clear that a larger $f^2/V$ leads to a larger frequency splitting ($2g$) and a larger linewidth difference ($2\Gamma$) between the split modes. For a spherical scatterer of radius $R$, $\alpha$ is given by $\alpha = 4\pi R^3 (n_p^2 - n_m^2) / (n_p^2 + 2n_m^2)$, where $n_p$ and $n_m$ denote the refractive indices of the scatterer and the surrounding medium, respectively[S3]. It is seen that $\alpha$ depends on both the scatter size and the contrast of refractive index between the scatterer and the medium surrounding the scatterer. In the simulations and the experiments presented in this study, we consider spherical scatterers for simplicity. However, the conclusions could be extended to scatterers of any shape.

When nanoparticles are deposited consecutively on a resonator surface, with each new particle entering the mode volume, the established SWMs are perturbed and redistributed such that the coupling strength between CW and CCW modes is maximized[S21]. As a result, one SWM experiences the maximum resonance shift whereas the other SWM goes through the minimum resonance shift, leading to a maximum frequency splitting between them. Therefore, the distributions of the two SWMs within the resonant cavity are modified by each arriving particle in the mode volume, and the resulting total frequency splitting is determined by all the particles on the resonator surface. Assuming there are $N$ particles continuously bind on a resonator, we denote the spatial phase difference between the first particle and the anti-node of one of the SWM as $\phi_N$, and the phase distance between the 1$^{st}$ and the $i^{th}$ particles as $\beta_i$. Then the frequency splitting $\delta_N^-$ and the linewidth difference $\rho_N^-$ of the two SWMs after adsorption of $N$ particles are written as[S21]

$$\delta_N^- = \left|\sum_{i=1}^{N} 2g_i \cos(2\phi_N - 2\beta_i)\right| \tag{S2}$$

$$\rho_N^- = \left|\sum_{i=1}^{N} 2\Gamma_i \cos(2\phi_N - 2\beta_i)\right| \tag{S3}$$

where $2g_i$ and $2\Gamma_i$ correspond to the frequency splitting and the linewidth difference if the $i^{th}$ particle is the only particle in the resonator mode volume. Maximization of the frequency splitting after adsorption of $N$ particles leads to

$$\tan(2\phi_N) = \frac{\sum_{i=1}^{N} g_i \sin(2\beta_i)}{\sum_{i=1}^{N} g_i \cos(2\beta_i)}. \tag{S4}$$

We can see from Eqs. (S2) and (S3) that frequency splitting and linewidth difference between two SWMs are characterized by the sum of the effect from each individual particle whose contribution is weighted by its position relative to the SWMs. This is explained by the fact that scatterer-induced coupling strength depends on the light intensity at the location of the scatterer which is position dependent for SWMs. With each particle entering the mode volume, the frequency splitting and linewidth difference change correspondingly.

Equation (S2), together with the expression for 2*g*, indicates that the amount of frequency splitting is related to the location of the particle within the mode volume of the WGM and with respect to other adsorbed particles on the resonator (i.e., $\delta_N^-$ is related to $f(\mathbf{r})$ and $\beta_i$). The total frequency splitting after adsorption of many particles carries the information of polarizability for each particle as well as their locations. In experiments it is difficult and technologically challenging to place each particle at a specific pre-determined position on the resonator. For the results presented in this work, we assume particles land on random locations on the resonator surface which is the case in our experiments. Then for each arriving particle, the amount of frequency splitting shows discrete jumps with different heights. Similar results are observed for particle sensing using single resonant mode, i.e. different frequency shifts in response to consecutively adsorbed particles in the mode volume of a resonator[S22].

Numerical simulations confirmed the dependence of discrete changes in frequency splitting on the location of particles in the WGM mode volume. The cross-sectional distribution of the fundamental WGM field shown in Fig. S1a implies that the optical field along the resonator surface is non-uniform, and thus light-matter interaction strength varies depending on the position of the particle on the resonator. Consequently, a single particle adsorbed on different locations in the mode volume induces different frequency splitting. Figure S1b depicts the results of Monte-Carlo simulations in which we calculate the frequency splitting of a WGM when individual Polystyrene (PS) nanoparticles of the same size are adsorbed onto the resonator one by one. With particles binding to random locations on the resonator surface, the resulting frequency splitting either increases or decreases with different step heights.

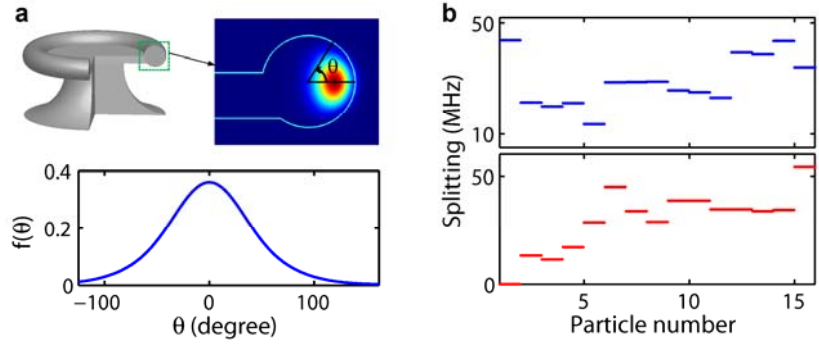

**Figure S1. a**, Upper panel: Illustration of a microtoroid and cross-sectional field distribution of the fundamental WGM. Lower panel: Normalized distribution of electrical field magnitude $f$ along resonator surface as a function of parameter $\theta$. **b**, Simulation results from Eqs. (S2) and (S4) depicting the frequency splitting as PS nanoparticles are continuously and randomly deposited on a microresonator. The upper and lower panels are obtained for two sets of simulations with nanoparticles placed at random positions on a resonator. Parameters used in simulations are: PS particles with radius $R = 50$ nm and refractive index $n_p = 1.59$; resonance wavelength $\lambda_c = 1550$ nm; surrounding medium with refractive index $n_s = 1.0$ (air); and a resonator with mode volume $V = 300$ μm$^3$. Each nanoparticle adsorption event leads to an upward or downward jump in the frequency splitting. Step height of each jump depends on the particle location in the mode volume.

Figures S2a and S2b depict the simulation results of single PS particle-induced $2g$ and $2\Gamma$ as a function of particle radius $R$ for two different resonator parameters $V$ and $f$. Figures S2c and S2d show the resonance frequency shift and linewidth broadening of the two orthogonal SWMs as PS particles of radius 30 nm are deposited consecutively and randomly onto the resonator surface. The red shift in resonance frequency is attributed to the larger refractive index of PS particles than that of air[S3]. It is seen that, small particles affect the resonance shift $2g$ more than the linewidth broadening $2\Gamma$. This can be explained by the $R^3$ and $R^6$ dependences of $2g$ and $2\Gamma$, respectively, on the particle radius $R$. In our experiments, particles are a few tens of nanometers, and therefore their effect on the linewidth broadening is significantly lower than their effect on the resonance shift / frequency splitting. In this study we focus on the detection and measurement of changes in frequency splitting.

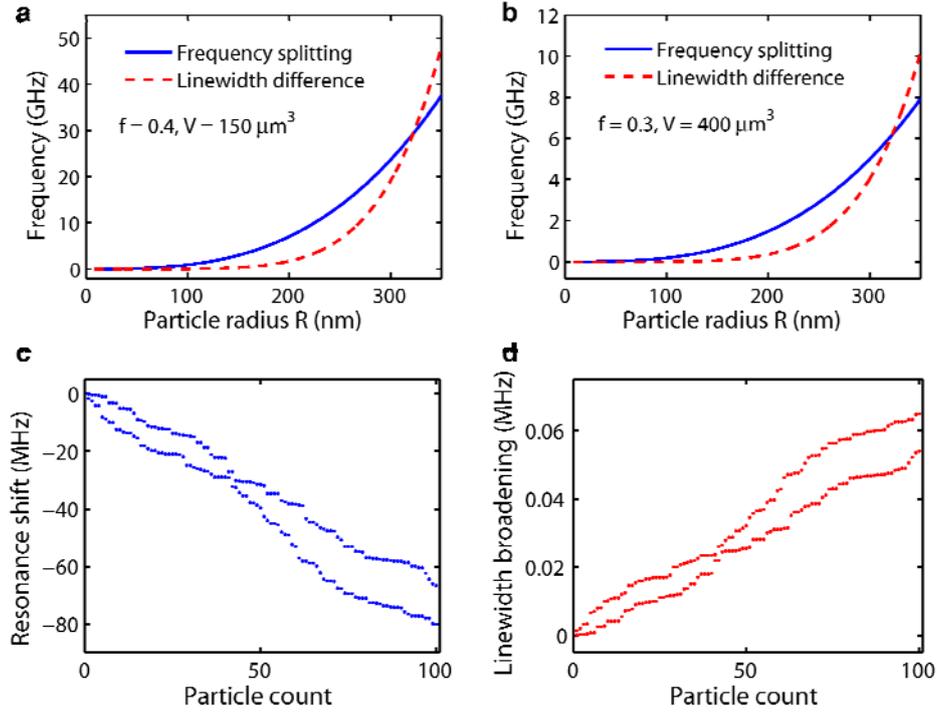

**Figure S2. a,b**, Calculation results of frequency splitting $2g$ and linewidth difference $2\Gamma$ induced by a single PS nanoparticle of different sizes binding on a resonator with different parameters of mode volume $V$ and normalized electrical field $f$ at the particle location, as labeled in the plot. Simulation results of the resonance frequency shift (**c**) and linewidth broadening (**d**) of two SWMs as PS particles are consecutively deposited on the resonator surface randomly. Parameters used in the calculations of (**c**) and (**d**) are: radius of PS particles $R = 30$ nm; maximum normalized field $f = 0.3$ (i.e., depending on the position of each nanoparticle on the resonator, $f$ varies in the range 0~0.3); and mode volume $V = 400$ μm$^3$. The results are obtained at resonance wavelength $\lambda_c = 1550$ nm.

The above discussions on mode splitting consider the optical mode supported in a resonant cavity regardless of whether the mode is excited by an external light source coupled into the cavity or is generated through stimulated emission from the gain medium embedded inside the cavity itself (microlaser). Thus, above discussions and conclusions are valid for both passive microresonators and microlasers.

**Effect of cavity size on the sensitivity of the scheme**

From the expression $2g = -\alpha f^2(r)\omega_c/V$, scatterer-induced frequency splitting depends on the field distribution function $f$ and the mode volume $V$, both of which are cavity-related parameters. For a particle of polarizability $\alpha$, the higher the value of $f^2(r)\omega_c/V$, the larger the splitting, and hence the better the detection sensitivity[S1]. This is confirmed with numerical simulations for single PS nanoparticle-induced $2g$ versus the particle size at various values of $f$ and $V$ (Fig. S3).

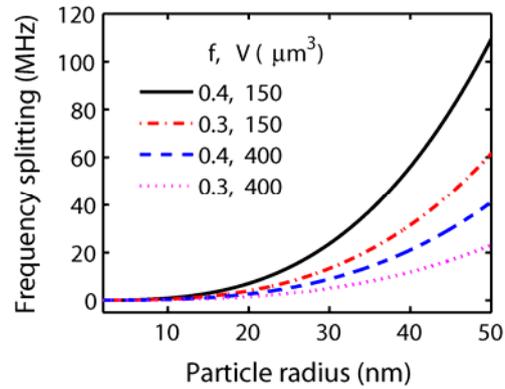

**Figure S3.** Dependence of a single PS nanoparticle-induced frequency splitting $2g$ on the particle radius $R$ for various values of normalized field $f$ and mode volume $V$. Simulations are done at resonance wavelength $\lambda_c = 1550$ nm.

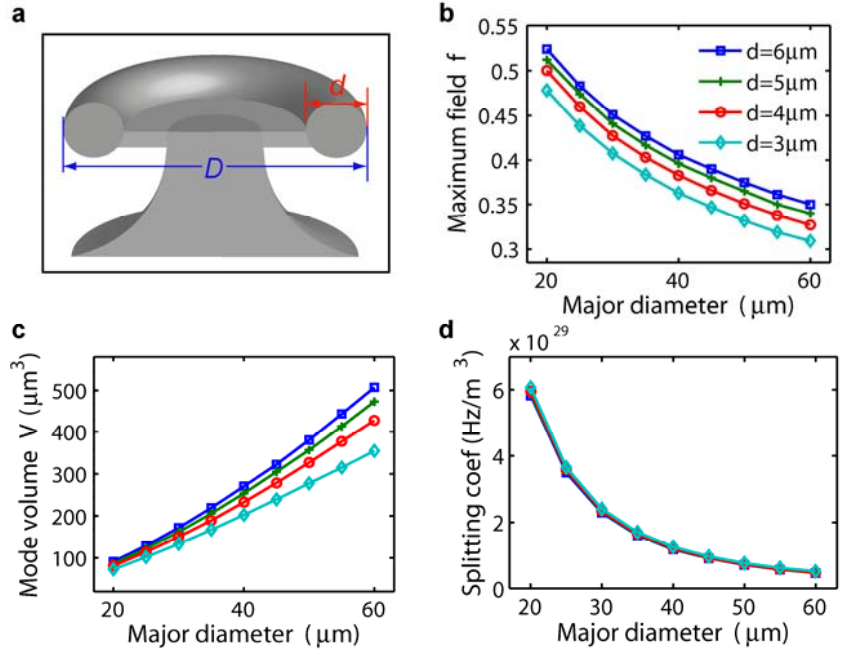

**Figure S4. a**, Illustration showing the major ($D$) and minor ($d$) diameters of a silica toroidal microresonator. Calculated maximum electric field magnitude $f$ at the outer boundary of the cavity (**b**), mode volume $V$ (**c**), and splitting coefficient (**d**) versus major diameter for minor diameters of $d$ = 6, 5, 4, and 3 μm. The plots are for TE-polarized fundamental WGM. The splitting coefficient in (**d**) is defined as $f^2\omega_c/V$ at resonance wavelength $\lambda_c$ = 1550 nm. The legends of data points in (**c**) and (**d**) are the same as those in (**b**).

Since both $f$ and $V$ are functions of physical dimension of the microcavity, we performed simulations using finite element method to characterize resonant modes in a microtoroidal resonator and to understand the dependence of $f$ and $V$ on its major and minor diameters (Fig. S4a). The field function $f$ is a position-dependent parameter and its value varies with the radial and azimuthal positions. In Fig. S4, for simplicity, we calculated the maximum field on the outer boundary of the resonator. Figure S4 reveals that when minor diameter is fixed, $f$ is smaller and $V$ is larger for the microtoroid with a larger major diameter, i.e., $f$ decreases whereas $V$ increases with increasing major diameter. On the other hand, for a fixed major diameter, both $f$ and $V$ increase with increasing minor diameter. This can be understood as follows. For a smaller resonant cavity, the spatial confinement is stronger and thus the optical mode is more compressed in the resonator leading to a smaller mode volume. On the other hand, the evanescent field leaking out into the surrounding medium is larger resulting in a higher field function on the surface of the resonator. Defining $f^2\omega_c/V$ as the splitting coefficient, and plotting it as a function of the major diameter, we see that splitting coefficient decreases with increasing major diameter of the resonator, while the effect of minor diameter is very small, if not negligible. These results imply that smaller microtoroids have better sensitivity than bigger ones.

In Fig. S5, we present two-dimensional mode distributions in the cross-section area of the ring for microtoroids with different major and minor diameters, together with the one-dimensional mode distributions along the periphery and the radial direction of the ring labeled with red dotted lines in Fig. S5a. In a microtoroidal resonator, the resonant mode is confined near the surface of the ring. The effects of physical dimension of the microtoroid on the resonant mode confinement as well as the field distributions are clearly seen in Fig. S5.

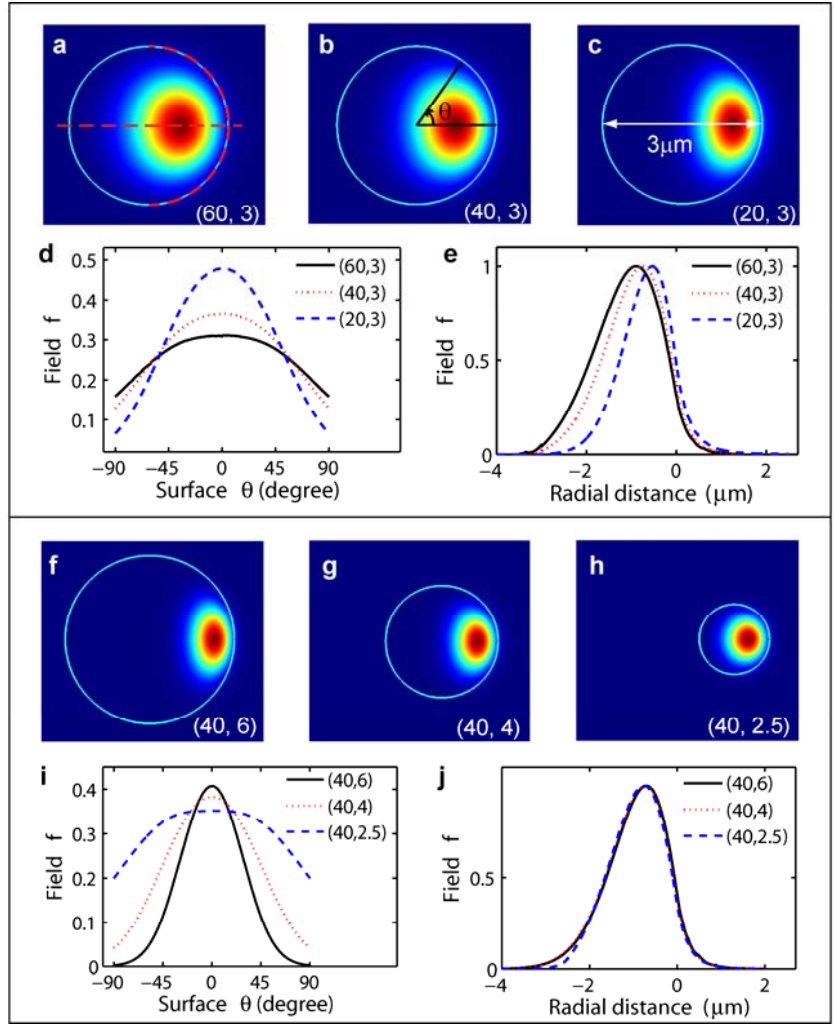

**Figure S5.** Profiles and distributions of electric field magnitude of TE-polarized fundamental modes in microtoroids of different major and/or minor diameters. Size of the microtoroid is denoted in a pair of parentheses as (major diameter, minor diameter) in micrometers marked in each plot. (**a**)-(**e**) depict the results for different major diameters and the same minor diameter. (**f**)-(**j**) show the distributions for the same major diameter and different minor diameters. (**a**)-(**c**) and (**f**)-(**h**) present the two-dimensional field distribution along the cross-sectional area of the ring. (**d**) and (**i**) present the normalized field distribution along the resonator surface, marked by the red dotted line along the outer boundary of the microtoroid in (**a**). (**e**) and (**j**) show the normalized field distribution along the central cross line of the ring as marked by the red dotted line in the horizontal direction in (**a**). In (**e**) and (**j**), the point $x = 0$ represents the boundary of the microtoroid.

We performed experiments by depositing PS nanoparticles of mean radius $R$ = 50 nm and coefficient of variation (CV: the ratio of standard deviation to mean) 5.1% one by one on two Erbium (Er)-doped toroidal microlasers with different diameters, and plotted the histograms of discrete changes in the beat frequency in Fig. S6. It shows that the standard deviation of discrete changes is larger for the smaller microcavity laser, implying that a smaller cavity is more likely to lead to a bigger change in the frequency splitting for the same perturbation and thus has better detection sensitivity.

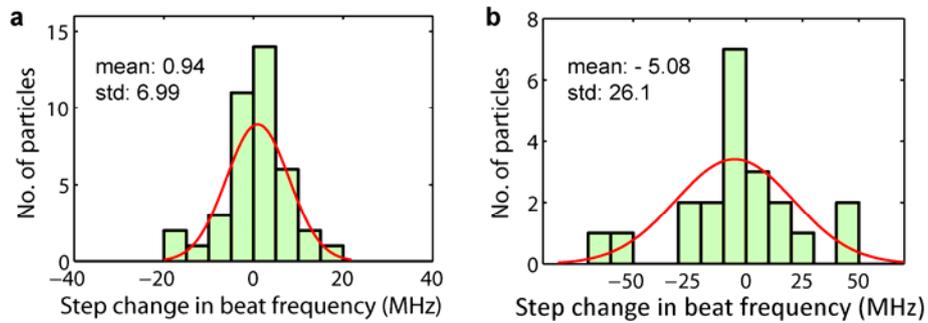

**Figure S6.** Histograms of measured discrete changes in the laser beat frequency as PS nanoparticles of radius $R$ = 50 nm are deposited on Er-doped microtoroids of diameter 42 μm (**a**) and 30 μm (**b**). The total number of detected particles is 40 for (**a**) and 21 for (**b**). Red solid lines in (**a**) and (**b**) are Gaussian fittings. The mean and standard deviation of the discrete changes in beat frequency are depicted in each plot with unit MHz.

**Noise level, detection limit, and detectable change in polarizability**

In our experiments, the smallest detectable particle size is determined by the fluctuation noise in the laser beat frequency. In order to quantify this noise, we recorded the beat frequency under the same condition without inducing any particles. Figure S7a shows the detected beat frequency with time. The noise level ±100 kHz implies that particle adsorption-induced splitting change within this noise level cannot be detected. The beat frequency noise has a Gaussian distribution as seen from the measured histogram in Fig. S7b. The noise can be attributed to various factors including (i) pump laser fluctuations, (ii) taper-cavity gap fluctuations, (iii) noise from photodetector, (iv) data processing noise, and (v) contaminants or Er-ion clusters inside the microlaser whose polarizabilities fluctuate due to the intracavity thermal fluctuations.

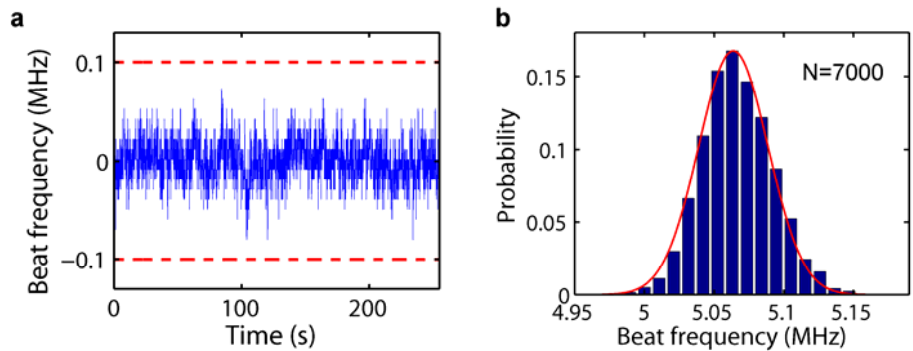

**Figure S7. a**, Noise level of the beat frequency fluctuation: noise is within ±100 kHz. This noise level sets the limit for the smallest detectable particle size in our experiments. **b**, A typical histogram showing the distribution of the beat frequency noise. The histogram is constructed from a total of 7,000 measurement points. Red solid line is a Gaussian fitting with mean 5.06 MHz and standard deviation 25 kHz.

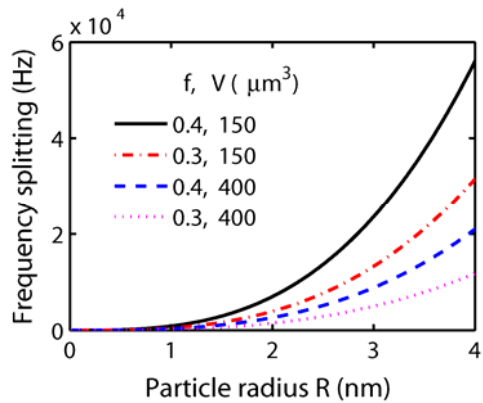

**Figure S8.** Simulation results of a single PS particle-induced frequency splitting versus particle radius for different values of $f$ and $V$. Simulations are done with resonance wavelength $\lambda_c$ = 1550 nm.

In Fig. S8, we depict the numerical simulation results showing the frequency splitting as a function of the PS particle radius for various values of $f$ and $V$. For cold cavity $Q$ in the range $10^6 \sim 10^7$, the ultimate detection limit for single nanoparticles using the microlaser-based scheme varies from less than 1 nm to a few nm.

Another important parameter in sensing applications is the minimum detectable change in the properties of the particle which leads to an observable change in the sensing signal. In other words, if the size $R$ or the refractive index $n$ of a nanoparticle is perturbed by some means, the question is whether the sensing signal (beat note frequency or frequency splitting) can catch such changes or not. In our detection scheme, such changes in an adsorbed particle on the resonator lead to changes in the laser beat frequency, and thus should be detected. If this is achieved, the method can be used to study the dynamical changes in particle shape, size, and refractive index. Figure S9 shows the detectable change $\Delta R$ in the radius $R$ of a PS nanoparticle for two different beat frequency noises. The minimum detectable change in radius decreases quadratically with $R$, and increases linearly with the beat frequency noise. Thus, in principle, the proposed scheme could be used as a sensitive measurement to monitor changes in the properties of a nanoparticle.

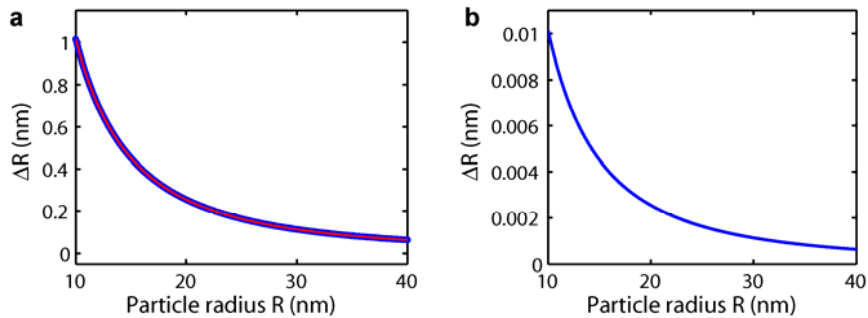

**Figure S9.** Simulation results showing the detectable change in the size of a PS particle given beat frequency noise of 100 kHz (**a**) and 1 kHz (**b**). Calculations are done with $\lambda_c = 1550$ nm, $f = 0.4$, and $V = 400$ μm$^3$. The red solid curve in (**a**) is a fitting of $1/R^2$.

**Experiments in air.** The detection of single nanoparticles and virions in air is done using Er-doped silica microtoroidal resonators[S23-24]. Figure S10 shows examples of the recorded frequency splitting, the beat note signal and the corresponding frequency spectra when gold nanoparticles are continuously deposited in the microlaser mode volume. Each particle binding event is signaled by a change in the frequency splitting reflected as a discrete change in the laser beat note frequency.

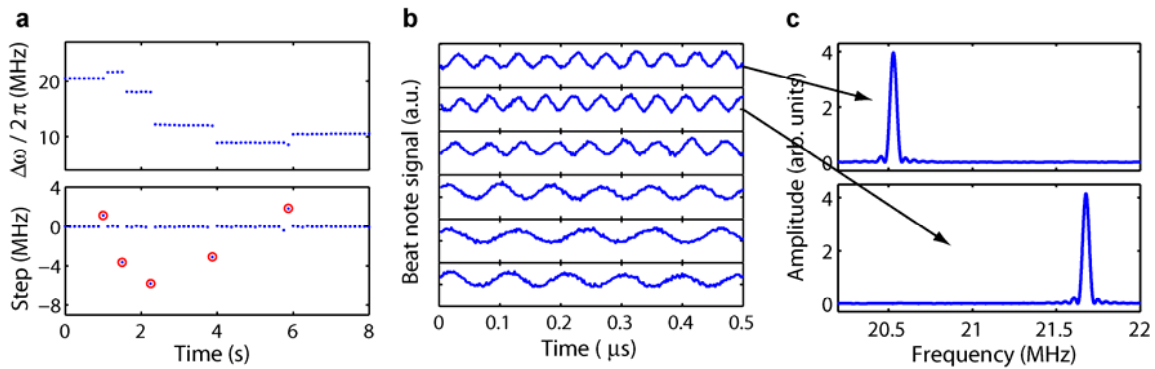

**Figure S10**. **a**, Recorded real-time beat frequency when gold nanoparticles (radius $R$ = 50 nm, CV < 8%) are continuously deposited onto the microlaser. Lower panel of (**a**) shows the frequency difference of the beat note between two consecutive detected data points. Five particle binding events marked with red circles are clearly seen from the discrete jumps. **b**, Beat note signals in each time segment in (**a**). **c**, FFT spectra of the first and second beat notes from top in (**b**) as marked by black arrows. Clear shift of the FFT frequency is observed corresponding to the first particle binding event in (**a**).

**Experiments in water.** We test the applicability of using a microlaser for single nanoparticle detection in aqueous environment. The experimental setup is similar to that used for experiments in air except that the fiber taper and the microlaser are immersed in a chamber filled with water. The Er-doped microtoroid laser cannot be used in water due to the high absorption of water in 1550 nm band. It degrades the resonator $Q$ factor to values less than $10^5$ which is not sufficient to generate laser emission with pump power at sub-mW level[S25]. Therefore, for water experiments we use Ytterbium (Yb)-doped microtoroids, which generate laser emission in 1040 nm band when pumped in 970 nm band. Water absorption in 1040 nm band is much lower than that in 1550 nm band giving rise to a higher cavity-$Q$ in water [S26]. The preparation of Yb-doped microtoroids follows the same procedure of preparing the Er-doped microtoroids except that the dopant is changed from $Er^{3+}$ to $Yb^{3+}$. The diameter of the Yb-doped microtoroids is around 100 μm. This big size helps reduce the radiation loss induced by the small refractive index contrast between silica and water. The measured $Q$ factor of the Yb-doped microtoroid in water is ~ $6\times10^5$ at 970 nm wavelength band. We observed laser emission in water with Yb ion concentration of $5\times10^{19}$ $cm^{-3}$, and the lasing spectrum is presented in Fig. 5a in the main text. In the same figure, we also included the changes in beat frequency as particles bind to the microlaser surface.

**Ensemble measurement of particle size by statistical analysis of discrete changes in beat frequency**

For random deposition of nanoparticles on a resonator, the frequency splitting either increases or decreases with different step heights. However, statistics of an ensemble measurement on the step changes in frequency splitting contains information of the particle size. For example, Fig. 11Sa shows the calculated histograms of changes in frequency splitting corresponding to 200 PS particles binding events for two different particle sizes. It is seen that larger particles result in a broader distribution.

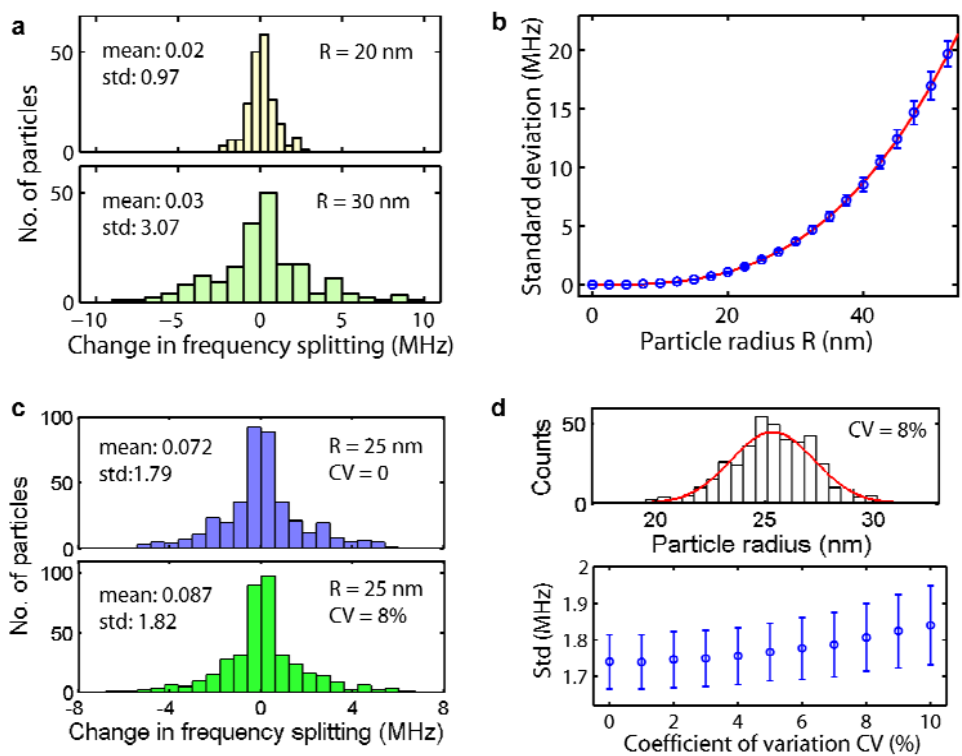

**Figure S11. a**, Histograms showing the distributions of changes in frequency splitting corresponding to deposition of 200 PS nanoparticles of radius $R$ = 20 nm and $R$ = 30 nm on a microresonator. The mean and standard deviation are depicted in the plot with unit MHz. It is seen that nanoparticles with smaller size lead to a narrower distribution. **b**, Standard deviation of changes in frequency splitting corresponding to binding of 200 PS nanoparticles with various sizes. Blue circles and error bars represent the mean $S_i$ and standard deviation $\sigma_i$ of the calculated values for repetition tests of 400 times. Red solid line is a polynomial fitting which linearly scales with $R^3$ (see text for details). **c**, Histograms of changes in frequency splitting for continuous adsorption of 400 PS nanoparticles of mean radius $R$ = 25 nm without (upper panel) and with (lower panel) size variation, assuming the particle radius has a Gaussian distribution with mean 25 nm and CV = 8%. **d**, Upper panel: distribution of particle radius. Red solid line shows the Gaussian fitting. Lower panel: calculated $S_i$ (circles) and $\sigma_i$ (error bars) as a function of CV of the PS particle radius. The particle radius has a Gaussian distribution with mean of 25 nm.

In order to understand the size dependence of the statistics of particle-induced changes in frequency splitting, we performed simulations using PS particles of different sizes. For each particle radius $R_i$ (i.e., $i = 1, \ldots, 22$), we deposited 200 PS particles one by one randomly in the mode volume of a resonator, recorded changes in the frequency splitting corresponding to all particle binding events, and calculated the standard deviation of these changes. We repeated this process $N = 400$ times for each particle size, and obtained a set of standard deviation $S_{ij}$ of the changes in frequency splitting where $j = 1, \ldots, N$. Then from the recorded set of $S_{ij}$, we calculated the mean $S_i = \langle S_{ij} \rangle$ and the standard deviation $\sigma_i = \sqrt{\frac{1}{N}\sum_{j=1}^{N}(S_{ij} - S_i)^2}$, and depicted them in Fig. S11b. A curve fitting to the numerically obtained data reveals a third-order relation between $S_i$ and $R_i$, that is, $S_i \propto R_i^3$. This originates from the linear relation between frequency splitting $2g$ and polarizability $\alpha$, and the third order relation between $\alpha$ and $R$ (i.e., $\alpha \propto R^3$). In experiments, the larger the measured particle number, the closer the obtained standard deviation to its statistical mean value $S_i$. We can construct a calibration standard by preparing a look-up table of $S_i$ through experiments using ensembles of a specific type of particles with known size $R_i$. This standard can be used to estimate the unknown polarizability of particles in an ensemble by plugging the standard deviation from the measurements into the calibration standard.

In Figs. S11a and S11b, we present the distributions of jump heights of the frequency splitting for nanoparticles of uniform size deposited randomly on a resonator. It is seen that as particle size increases, the standard deviation of the splitting changes induced by particle attachment increases, implying that the standard deviation of these changes carries information of particle size. In reality, however, it is difficult to have ensembles of identical particles, and there is always variation in the size of manufactured particles even in the same batch. That is why average size and CV are used to characterize the particle ensembles. In Figs. S11c and S11d, we take into account the size variation of the adsorbed nanoparticles and show their influence on the measurements. Figure S11c displays the effect of CV on the histograms of splitting changes for 400 PS nanoparticles continuously and randomly deposited onto a resonator. For comparison, we considered two ensembles of PS particles with mean radius of $R = 25$ nm. In the first ensemble, all particles have the same size with CV equals to 0, whereas in the second ensemble CV is 8% and a Gaussian size distribution is assumed (upper panel in Fig. S11d). It is seen that the histograms of splitting changes do not show significant difference for the two cases. The maximum change in frequency splitting becomes larger for the ensemble with CV = 8% (lower panel in Fig. S11c). This is because of the fact that CV of 8% leads to a size distribution in the range of 18.9-30.2 nm with mean radius 25 nm (upper panel in Fig. S11d). The maximum change in frequency splitting is obtained for particles with radii close to the maximum value of 30.2 nm. Simulation results in Fig. S11d reveal that the distribution of particle size leads to a broader distribution in the histogram of splitting changes. For example, the standard deviation of splitting changes increases from 1.74 MHz for CV = 0 to 1.84 MHz for CV = 10%, corresponding to a change of 5.7%. Since the estimated size has a cubic root dependence on the frequency splitting, such small change in the standard deviation will not significantly affect the estimation of the mean particle size.

We performed experiments with ensembles of gold particles of $R_1 = 15$ nm (mean diameter 30.3 nm, CV< 8%) and $R_2 = 25$ nm (mean diameter 49.7 nm, CV < 8%) to confirm the size estimation with the above approach. The mean diameter and CV are provided by the manufacturer. After collecting statistically significant number of discrete changes in the beat frequency, we constructed histograms of the steps for the tested particles as shown in Fig. S12a. Gaussian fittings of the histograms are presented in the plot. The histogram for $R_2 = 25$ nm particles has a wider distribution than that for $R_1 = 15$ nm particles which is consistent with the above simulation results. However, we cannot extract the size information by simply applying the relation $S_i \propto R_i^3$. This is because in the experiments, we can only detect the steps in beat frequency that are larger than the noise level, that is, if the particle-induced change in beat frequency is less than the noise level, it will go undetected. Thus, binding events which introduce very small changes in beat frequency can not be detected and the information of such events is lost. This is observed as the gap around 0 in histograms depicted in Fig. 3c and Fig. S12a. To solve the problem, we select a threshold step $\delta$, and define $S$ as the standard deviation of the measured steps that are greater than $\delta$. Thus $S$ is a function of $\delta$. From the collected data, we calculated a dimensionless parameter *"weighted standard deviation $\mu$"*, defined as $\mu_1 = S_1/\delta$ for particles of $R_1$ and $\mu_2 = S_2/\delta$ for particles of $R_2$, and plotted them as a function of the threshold $\delta$ in Fig. S12b. Denoting $\delta$ which lead to $\mu_1 = \mu_2$ for the two ensembles as $\delta_1$ and $\delta_2$, we found that $(\delta_1/\delta_2) \propto (R_1/R_2)^3$. Therefore, from the ratio $\delta_1/\delta_2$, we will get the size relation between $R_1$ and $R_2$. Using this ratio $\delta_1/\delta_2$ as a correction term (i.e., scale the x-axis of the curve for $R_2$ with $\delta_1/\delta_2$), we re-plotted the $\mu$ versus $\delta$ curves for the two ensembles and depicted them in the inset of Fig. S12b. There is a good overlap between the two curves after correction. This is expected since the step changes in beat frequency for particles of $R_1$ and $R_2$ have the same type of distribution. This analysis suggests that we can calibrate the detection system using the measurement results performed on an ensemble of particles with known refractive index and known size, say $R_1$, and then estimate the unknown size $R_2$ of particles having the same refractive index with the calibration particles. If only the polarizability $\alpha_1$ is known, the same procedure will yield a relation between the polarizabilities of the reference and the unknown ensembles, i.e., $(\delta_1/\delta_2) \propto (\alpha_1/\alpha_2)$.

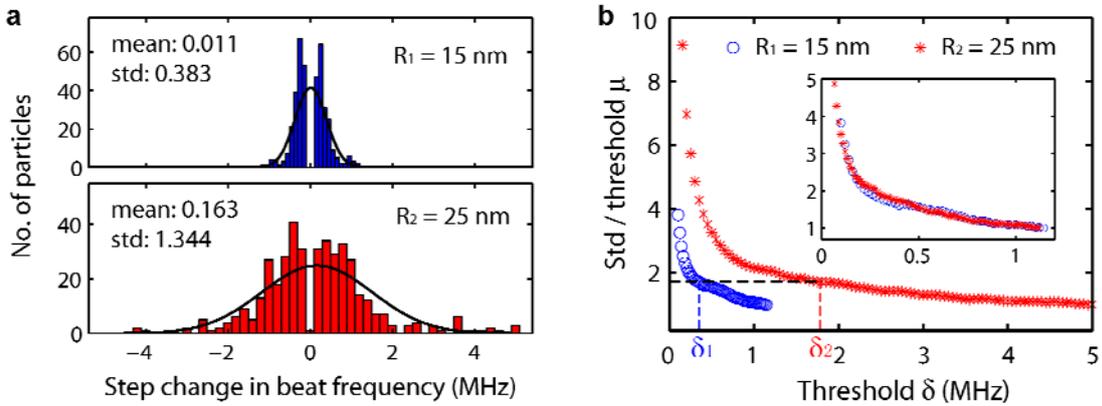

**Figure S12. a**, Measured histograms of the discrete jumps in beat frequency for two different gold particle sizes. Black solid curves are Gaussian fittings. The mean and standard deviation presented in the plot have unit MHz. **b**, Weighted standard deviation $\mu$ of splitting steps versus the threshold $\delta$ obtained from ensemble measurements of gold nanoparticles with two different sizes of $R_1$ = 15 nm (blue circles) and $R_2$ = 25 nm (red stars). Inset: The same curves plotted after multiplying the x-axis of the curve for $R_2$ with the correction term $\delta_1/\delta_2 = 0.22$, which is close to the ratio $(R_1/R_2)^3 = 0.23$ (see text for details).

It is worth noting that, for ensemble measurement, the maximum magnitude of the discrete changes in frequency splitting is also proportional to $R^3$, and could be used to estimate the particle size. However, since there are few particle-binding events near the maximum jump magnitude (Figs. S11a and S12a), in reality, any dust or contaminant induced splitting change might shift the maximum jump magnitude to a higher value and thus induce an estimation error. Therefore, the estimation of particle size from the maximum change in frequency splitting does not make full use of the detected data information and will introduce large error.